\documentclass[twocolumn,showpacs,preprintnumbers,amsmath,amssymb]{revtex4}
\usepackage{graphicx}
\usepackage{dcolumn}
\usepackage{bm}
\begin{document}
\title{Enhancement of Compton Scattering by an Effective Coupling Constant}
\author{Bernardo Barbiellini}
\affiliation{Department of Physics, Northeastern University, Boston, Massachusetts 02115, USA}
\author{Piero Nicolini}
\affiliation{Frankfurt Institute for Advanced Studies (FIAS),
Institut f\"ur Theoretische Physik, Johann Wolfgang Goethe-Universit\"at, 
Ruth-Moufang-Strasse 1, 60438 Frankfurt am Main, Germany}

\begin{abstract} 
A robust thermodynamic argument shows that 
a small reduction of the effective coupling constant $\alpha$ 
of QED greatly enhances the 
low energy Compton scattering cross section and that 
the Thomson scattering 
length is connected to a fundamental scale $\lambda$.
A discussion provides a possible quantum interpretation 
of this enormous sensitivity 
to changes in the effective coupling constant $\alpha$. 
\end{abstract}
\maketitle
\section{Introduction}
The process of the energy interchange between radiation and matter provided by Compton scattering is relevant in many areas of physics. 
For example, in cosmology it keeps the matter at the same temperature as radiation \cite{longair}. 
Compton scattering is also a unique spectroscopy for condensed matter physics, which has acquired greater importance with the advent of 
modern synchrotron sources \cite{PT_isaacs, cooper, schuelke}.  For instance, it has been used to extract information about wave functions 
of valence electrons in a variety of systems ranging from ice \cite{isaacs, nygard} and water \cite{sit} to alloys \cite{alli} 
and correlated electron systems \cite{bba09}. Moreover, Compton scattering can potentially help delineate confinements \cite{saniz1} 
and spin polarization effects \cite{saniz2}
in nanoparticles.

The Compton scattering cross section strength is 
determined by the classical electron radius, 
also known as the Thomson scattering length, 
\begin{equation}
r_0=\frac{e^2}{4 \pi \epsilon m c^2}\approx 2.82\times10^{-13} \mbox{cm}~,
\end{equation}
where $e$ is the electron charge, $m$ is the electron mass, $c$ is the speed 
of light and $\epsilon$ is the dielectric constant.  
Unfortunately, the small size of $r_0$ 
makes Compton experiments  
in condensed matter systems difficult. 
This is why only a few experiments have been done,
even with the best synchrotron sources. 
The classical proton radius is even smaller by a factor 
of $M/m\approx 1863$, 
where $M$ is the proton mass.
Therefore, nuclei are practically invisible in X-ray Compton scattering experiments.

In 1952, Max Born suggested that the electronic radius $r_0$ is 
connected to an absolute length scale $\lambda$ \cite{born0}. 
Thus, if the electromagnetic interaction strength is modified,
$\lambda$  must change as well.  Understanding this variation
could enable us to enhance the Compton scattering cross sections by 
{\em engineering} an effective quantum electro-dynamics 
(QED) interaction. The effective coupling constant
\begin{equation}
\alpha=\frac{e^2}{4 \pi \epsilon \hbar c}~,
\end{equation}
can be modified through the dielectric response $\epsilon$,
for instance, if the incident photon energy is tuned near to 
the binding energy of a deep core electron level in 
certain materials.

This work shows that the Compton 
cross section can depend strongly 
on the effective coupling constant  
$\alpha$ and that a reduction of $\alpha$ as small as
$1 \%$ may lead to an increase in the cross section 
by a factor of $4$.
Moreover, the present results connect $r_0$ 
to a fundamental length $\lambda$ and
thus are consistent with the old hypothesis by Born.

\begin{figure}
\begin{center}
\includegraphics[width=\hsize]{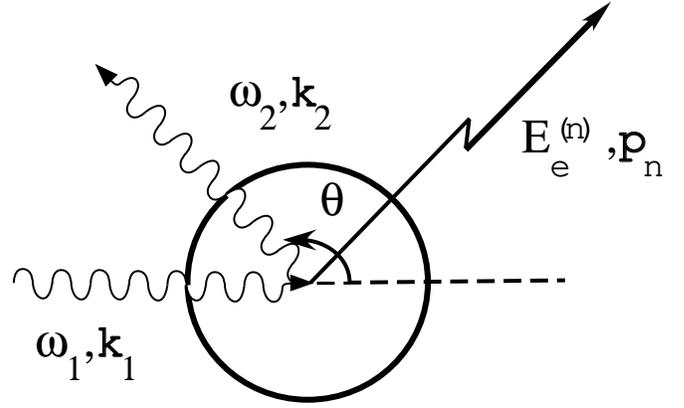}
\end{center}
\caption{Schematic diagram of the elementary scattering event 
involved in the Compton scattering process. 
The incoming photon scatters from the target to produce an 
outgoing photon and an electron and leaves the singly ionized target.}
\label{fig1}
\end{figure}

The triple-differential scattering cross section for the process 
shown in Fig.~\ref{fig1}, which is the elementary step underlying Compton scattering, 
is given by \cite{Kaplan,bba03}
\begin{eqnarray} 
\frac{d^{3}\sigma^{(n)}}{d\omega_{2}d\Omega_{2}d\Omega_{e}}=r_{0}^{2}
\frac{\omega_2}{\omega_1}(1+\cos^{2}\theta)\nonumber\\
\times |g_{n}({\bf q})|^{2}
\delta(\omega_{1}-\omega_{2}-E_{b}^{(n)}-\frac{p_{n}^{2}}{2m})~,
\label{eqcross}
\end{eqnarray} 
where $\theta$ is the scattering angle, $g_{n}({\mathbf{q}})$ 
is the Fourier transform of the occupied 
Dyson orbital $g_n({\bf r})$ with binding energy 
$E_b^{(n)}$, ${\bf q}$ is the momentum transferred to the final system, 
and $\omega_{1}$ and $\omega_{2}$ 
are, respectively, the energies of the photon before and 
after the collision. The ejected 
electron state is usually approximated by a plane wave with 
momentum ${\bf p}_n$ and energy $E_{e}^{(n)}=p_{n}^{2}/(2m)$
if $E_b^{(n)} \ll E_e^{(n)}$.  
In this regime, Compton scattering 
is a unique window on the electronic structure of matter because
in contrast with most structural analysis techniques which can only 
deliver information on the total electron densities, 
this spectroscopy allows direct measurements in momentum 
space of the electron density associated with a single ionization channel 
(i.e. a Dyson orbital in a one-electron picture).
In the low-energy limit (i.e. $\omega_1 \ll mc^2$), Thirring \cite{thirring} 
has shown that the Compton scattering cross section with 
all radiative corrections reduces in 
the non-relativistic expression given by Eq. (\ref{eqcross}). 
The only effect of the vacuum or the medium  
is to renormalize the Thomson scattering length $r_0$.
The {\em Thirring theorem} is a consequence of Lorentz 
and gauge invariance \cite{jauch,haeringen}.

\section{thermodynamic argument}
We now turn to a general thermodynamic argument
in order to derive how the electron
volume $V=4 \pi r_0^3/3$ depends on the 
effective coupling constant $\alpha$. 
Since the classical electron radius $r_0$ 
is the length at which QED renormalization effects 
become important, our argument
must be consistent with differential
equations of the renormalization group
\cite{bogoliubov}. 
Thermodynamics is widely considered as a 
fundamental theory, 
since it has a universal applicability 
\cite{Verlinde:2010hp,Nicolini:2010nb}. 
Indeed it does not need any modification due to 
either relativity or quantum theory \cite{paddy}. 
The first law of thermodynamics gives the variation 
of internal energy
\begin{equation}
dE =T dS -P dV + mc^2d \alpha~,
\label{eqde}
\end{equation}
where 
$T$ is the temperature,
$S$ is the entropy 
and $P=-E_s/V$ is  
a pressure imposed by a fictitious piston 
on the volume $V$ 
in order to set the units scale for a given 
$\alpha$ \cite{piston}.
Thus, the energy scale is characterized by
$E_s =\alpha^x~ mc^2$, where $x$ represents a
positive integer exponent to 
be determined.
The negative sign of the pressure $P$ is explained
by the fact that the electromagnetic vacuum fluctuation
(i.e., the Casimir effect) tries to
pull the piston back  into the system.
Similar inward pressures  are produced
by  cosmological constants \cite{wright}.
The third term in Eq.~(\ref{eqde}) 
is similar to a chemical potential 
term, since the number of virtual photons is proportional 
to the effective coupling constant $\alpha$. 
Thus, we are assuming that the electron mass $m$ 
determines the chemical potential of the virtual photons
and that it is generated by the 
Coulomb field of the electron.
In adiabatic conditions, the term $TdS$ vanishes.  
Moreover, at equilibrium, $dE=0$, thus the 
renormalization group $\beta$ function 
\cite{bogoliubov}
deduced from Eq.~(\ref{eqde})
is given by
\begin{equation}
\beta(\alpha)= r \frac{d\alpha}{d r}=-3\alpha^x~.
\end{equation}
The solutions for $x=0,1,2$ show that the electron 
localizes (i.e., $r_{0}$ becomes small) when the interaction
strength increases.
When $x=0$, the radius scales as
\begin{equation}
r_{0} = r_{max}  \exp ( - \alpha /3),
\end{equation}
and has a maximal finite size $r_{max}$
corresponding at $\alpha=0$
while for $x=1$, the scaling is
\begin{equation}
r_{0} = \frac{\lambda_1}{\alpha^{1/3}},
\end{equation}
where $\lambda_1$ is radius corresponding 
at $\alpha=1$.
The exponent $x=2$ is consistent with 
the QED $\beta$ function 
\cite{bogoliubov}.
The Born hypothesis
is also verified 
when $x=2$, since 
the corresponding solution admits a minimal
length $\lambda$ different from zero.
In this case, the Thomson scattering length
depends on $1/\alpha$ by an exponential function
\begin{equation}
r_0=\lambda\exp \left(\frac{1}{3\alpha}\right)~,
\label{eqr0}
\end{equation}
where $\lambda$ is a certain small length to 
be determined.
Moreover, the corresponding pressure 
$P=\alpha^2~ mc^2$ sets 
the atomic energy units. 
In fact, the atomic units are as natural as the fundamental
Planck units \cite{manin}: their ratios to 
the fundamental units 
can be explained 
within our present argument 
connecting the Thomson scattering 
length to the fundamental scale.
Interestingly, the volume renormalization 
factor $Q(\alpha)$ 
is $\exp(1/\alpha)$ for $x=2$. This term 
is similar to the Boltzmann distribution 
in statistical mechanics (where $\alpha$
plays the role of an effective temperature). 

\begin{figure}
\begin{center}
\includegraphics[width=\hsize]{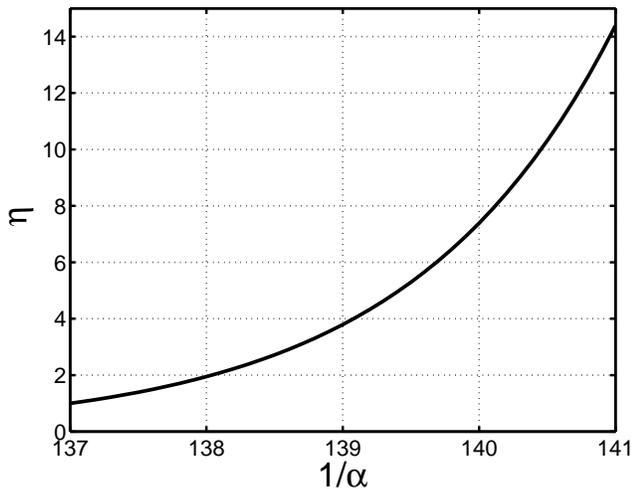}
\end{center}
\caption{Cross section enhancement $\eta$ as a function of 
the inverse of the effective coupling constant
$\alpha$. Both $\eta$ and $\alpha$ are pure numbers 
(units are dimensionless).}
\label{fig2}
\end{figure}
The cross section enhancement defined by
\begin{equation}
\eta= \left[ \frac{Q(\alpha)}{Q(1/137)} \right]^{2/3}
\end{equation}
is shown in Fig.~\ref{fig2} for the case $x=2$:
a reduction of $\alpha$ by few percents induces a huge 
increase in $\eta$. Therefore, cross section 
enhancements obtained 
by tuning the incident photon energy near 
the binding energy of 
a deep core electron level 
can be described by the behavior for 
$x=2$ while the cases $x=0,1$ give negligible
cross section enhancements for small variations of $\alpha$.
The trend of $\eta$ illustrated in Fig.~\ref{fig2} can be produced
by a change $\Delta \epsilon$ of the dielectric response 
near an absorption edge.  

\section{Proposed experiment}
Standard  
inelastic X-ray scattering experiments without
the measurement of the kinematics of the outgoing (recoil) electron
contain many other processes in addition to the elementary 
scattering event of Fig.~\ref{fig1}.
Therefore, coincidence $(\gamma, e \gamma)$ experiments \cite{Kaplan}
are needed in order to separate the X-ray Compton scattering 
with nearly free electrons from complicated processes.
Some $(\gamma, e \gamma)$ spectrometers are already 
available for hard X-rays \cite{itou}.
Unfortunately, standard $(\gamma, e \gamma)$  experiments 
can be tremendously challenging.
Instead, one could use a soft-x-ray fluorescence spectrometer 
by Carlisle {\em et al.}  \cite{carlisle}. 
By tuning the incident photon at the $K$ edge of graphite,
enhancement effects of the total cross section have been already observed. 
A coincidence measurement detecting the electrons escaping from the sample 
can then be used to separate Compton from other types of inelastic scattering. 
In this much simpler
setup multiple scattering 
of the electrons in the sample
are not an impediment to extracting the Compton contribution. 

\begin{figure}
\begin{center}
\includegraphics[width=\hsize]{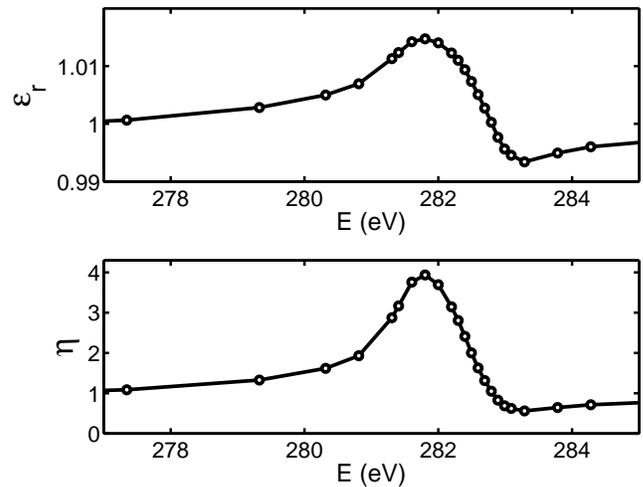}
\end{center}
\caption{
Top: $\epsilon_r$ 
(ratio of the real part of the 
dielectric function $\epsilon$ 
and the vacuum permittivity $\epsilon_0$) 
near the $K$ edge of graphite. 
$E$ is the incident photon energy.
The points are from Draine \cite{draine}. 
Symbol size is representative 
of error bars. 
Bottom: corresponding cross section 
enhancement $\eta$.}
\label{fig3}
\end{figure}

Realistic dielectric data for graphite provided by Draine 
\cite{draine} 
illustrate how tuning the 
incident photon energy near the binding energy of the $K$ core level
changes the dielectric response and 
thus the effective coupling constant for the valence electrons. 
When $x=2$, a Compton cross-section enhancement $\eta$ 
of almost a factor $4$ is predicted in graphite 
by using  Draine's dielectric data as shown in Fig.~\ref{fig3}. 
We note that a similar variation of the dielectric function for
diamond has been previously reported 
by Nithianandam and Rife \cite{rife}.
Besides, a calculation based on the 
finite difference method for near-edge structure
(FDMNES) \cite{joly}
agrees with the dielectric data of Draine.
FDMNES shows that the
anomalous scattering factor near the $K$ edge of graphite
becomes greater in amplitude than the number of electrons, causing 
the real part of $\epsilon/\epsilon_0$ to be greater than unity
($\epsilon_0$ is the vacuum permittivity).

\section{Discussion and concluding remarks}
Next, we justify a value for $\lambda$. 
According to Veneziano \cite{veneziano}, 
a consistent quantum gravitational theory should 
obey the Born principle of reciprocity \cite{born}, 
a symmetry law under the interchange of space-time coordinates and 
the energy-momentum coordinates, 
which naturally leads to harmonic oscillators and to the normal 
modes of vibrating strings. 
In such theory it is natural 
to take as the action quantum the square of the Planck length \cite{planck}
\begin{equation}
\lambda=\ell_P=\sqrt{  \frac{h G}{c^3}}
\approx 4.05 \times 10^{-33} \mbox{cm}~,
\end{equation}
where $h$ is the Planck constant and $G$ is the gravitational constant. 
Indeed, by using Eq.~(\ref{eqr0}) with the Planck length $\lambda$ 
and $\alpha=1/137.03604$, the calculated Thomson scattering length
is $r_0=2.79\times10^{-13} \mbox{cm}$, which differs about $1\%$ from its exact value. 
Minor renormalization effects of the gravitational constant
$G$ could improve the agreement \cite{hamber}.

Finally, we could also reverse the logic. 
In our treatment the length $\lambda$ is not fixed {\em a priori}. 
Therefore, we can use data from X-rays experiments in graphite
to get information about the size of $\lambda$. 
This would strongly vivify a big portion of the existing literature in quantum gravity for which the presence 
of an effective minimal length is assumed to describe the discretization of a quantum space-time. 
Presently, in the absence of any experimental signature for quantum gravity, such a minimal length is generically set between 
the electroweak scale of $\sim 10^{-16}$ cm and the Planck length. 
As a result we are opening the door to the possibility of determining an extreme energy effect
with sophisticated low energy experiments.
In addition since preliminary data seem to support the 
idea that $\lambda =\ell_P$ up to 1 $\%$, 
we can get more stringent constraints about the extension of 
the conjectured additional spatial dimensions with respect to what we currently know from 
the observed short scale deviations of Newton's law \cite{Nicolini:2010nb,newton}.

In conclusion, we suggest that the low energy Compton 
cross section for the valence (i.e., nearly free) 
electrons of graphite can be described within the framework of 
the Thirring theorem, implying that the only effect of 
the medium is to renormalize the Thompson scattering length $r_0$.
Besides, a general thermodynamic argument shows that 
the Compton scattering cross section grows exponentially
if the effective coupling constant $\alpha$ decreases. 
In particular, a striking enhancement is predicted when the 
incident photon energy is tuned near the binding energy of the $K$ 
core level of graphite. The present enhancement effect 
is also consistent with the QED renormalization group.

\begin{acknowledgments}
We are grateful to A. Widom, P.M. Platzman, G. Barbiellini, 
B.T. Draine, U. Amaldi, B. Tiburzi, P. Nath and D. Wood
for useful discussions. B.B. is supported by the 
US Department of Energy, Office of Science, 
Basic Energy Sciences contracts DE-FG02-07ER46352 
and DE-FG02-08ER46540 (CMSN) and benefited from the allocation of 
computer time at NERSC and Northeastern University Advanced Scientific Computation Center (NU-ASCC). 
P.N. is supported by the Helmholtz International Center for FAIR 
within the framework of the LOEWE program 
(Landesoffensive zur Entwicklung Wissenschaftlich-\"{O}konomischer
Exzellenz) launched by the State of Hesse.
\end{acknowledgments}


\begin{thebibliography}{99}
\bibitem{planck} In Ref.~\cite{veneziano} the Planck length is smaller and
it is given by $\lambda=\sqrt{ \hbar G/c^3}\approx 1.6 \times 10^{-33} \mbox{cm}$.
\end{thebibliography}
\end{document}